\documentclass{svproc}
\usepackage{url}

\usepackage{amsmath}
\usepackage{caption}
\usepackage{bbm}
\usepackage{subcaption}
\usepackage{graphicx}
\usepackage{placeins} 
\usepackage{amssymb}
\usepackage[dvipsnames]{xcolor}
\usepackage{pifont}

\begin{document}

\mainmatter              
\title{Longitudinal Modularity,\\ a Modularity for Link Streams}
\titlerunning{A Modularity for Link Streams}  
%
\author{Victor Brabant\inst{1} \and Yasaman Asgari\inst{2}
\and Pierre Borgnat\inst{3}
\and Angela Bonifati\inst{1}  \and Rémy Cazabet\inst{1}}
\authorrunning{Victor Brabant et al.} 
%
\tocauthor{Ivar Ekeland, Roger Temam, Jeffrey Dean, David Grove,
Craig Chambers, Kim B. Bruce, and Elisa Bertino}
\institute{
1. UCBL, CNRS, INSA Lyon, LIRIS, UMR5205, F-69622 Villeurbanne, France \\
2. Department of Mathematical Modeling and Machine Learning, Digital Society Initiative (DSI), University of Zurich, CH-8057 Zurich, Switzerland\\
3. CNRS, Univ de Lyon, ENSL, Laboratoire de Physique, F-69342 Lyon, France \\
\email{victorbrabant@gmail.com}}

\maketitle              

\begin{abstract} 
Temporal networks are commonly used to model real-life phenomena. When these phenomena represent interactions and are captured at a fine-grained temporal resolution, they are modeled as link streams. Community detection is an essential network analysis task. Although many methods exist for static networks, and some methods have been developed for temporal networks represented as sequences of snapshots, few works can handle link streams. This article introduces the first adaptation of the well-known Modularity quality function to link streams. Unlike existing methods, it is independent of the time scale of analysis. After introducing the quality function, and its relation to existing static and dynamic definitions of Modularity, we show experimentally its relevance for dynamic community evaluation.
\end{abstract}

\section{Introduction}

Complex networks are powerful tools for modeling real-life phenomena, such as social interactions, economic transactions, and biological interactions. A classic problem in this field is discovering the community structure of a given network. While there is an extensive body of literature addressing this challenge \cite{fortunato2016community}, most of it focuses on static networks. Temporal or dynamic networks are frequently found when representing real-life phenomena, and discovering communities in such data represent a challenge of its own. Various methods have already been proposed to handle temporal networks \cite{rossetti2018community}, however many of them are only able to work on slowly evolving graphs \cite{cazabet2020evaluating}, i.e., graphs that can be represented as sequences of well-behaved static networks.
Many real-world phenomena are instead composed of a flow of interactions occurring at a fast scale, that cannot be interpreted as a usual static network at any single point in time. Confronted to such data, the usual approach in the literature is to aggregate the data into sequences of snapshots, using a time-window. However, this approach raises a number of difficulties, such as the choice of an appropriate time-window, and the loss of temporal information resulting from the aggregation. Instead, such data is better modeled as \textit{stream graphs}, or \textit{link streams} \cite{latapy2018stream}, frameworks designed to directly manipulate such fine-grained temporal data, without any unnecessary aggregation. Many common network analysis problems have been redefined and adapted to the link stream case, such as clustering coefficient, connected components, or shortest paths\cite{latapy2018stream}, cliques as $\Delta$-cliques\cite{viard2016computing}, random walk centralities \cite{beres2018temporal}, etc. 

A few methods in the literature have proposed to detect communities in link streams, however those methods are
limited to detect some specific subcase of partitions, such as non-evolving communities \cite{matias2018semiparametric} or having only one step of evolution and requiring a temporal scale of analysis\cite{bovet2022flow}.
Instead, we propose in this work to adapt a well-known quality function for community detection in static networks, the Modularity. Although an adaptation of Modularity for sequences of snapshots exists\cite{Mucha_2010}, it cannot work on link streams. 
In this article, we propose the first generalization of Modularity for link streams, that we name Longitudinal Modularity (L-Modularity). 

Section \ref{sec:definitions} defines link streams and introduce notations we will use throughout the paper. It also defines the type of dynamic communities we are considering. Section \ref{sec:SOTA} provides an overview of related works and their limits. Section \ref{sec:longmod} introduces the definition of the Longitudinal Modularity we propose. Section \ref{sec:propdyncom} introduces a set of properties one should expect a good temporal community definition to respect, and show that L-Modularity does indeed respect those properties. Section \ref{sec:experimentation} demonstrate through experimentations the relevance of our quality function. Finally, we conclude in Section \ref{sec:discussion}.

\section{Definitions: Link Stream and Temporal Community}\label{sec:definitions}

Temporal networks are known under many names and formalisms, such as temporal networks, dynamic networks, time varying networks, etc.
Similarly, various concepts of temporal communities are used in the literature. In this section, we clarify the definition of link streams, and what makes this type of representation relevant to study, and similarly for the notion of temporal community.

\subsection{Link Streams}
\label{sec:linkstream}
\subsubsection{Slowly Evolving Graphs and Link Streams}

Although many formalisms exist to represent temporal networks, in this article we make a fundamental distinction between two distinct types: slowly evolving graphs (SEG) and Link streams. SEG \cite{cazabet2020evaluating} correspond to networks that can be seen as series of static graphs, or as a network evolving \textit{edge by edge}, while remaining at every time a conventional, well-behaved network, that can be studied with the tools of network science. Among networks being typical SEG, one can mention friendship in social networks, or yearly snapshots composed by aggregating interactions among users of an online platform. Multiple community detection methods have been developed to deal with such networks \cite{rossetti2018community}, e.g., those compared experimentally in \cite{cazabet2020evaluating}. 

SEGs are opposed to link streams, that correspond to the original form of many real data, in which interactions are simply collected as events $(uv,t)$, corresponding to an \textit{interaction} between nodes $uv$ at time $t$, e.g., a physical interaction, an instantaneous message by phone or on a social platform, etc. In such data, there is no well-behaved network at each point in time, thus the usual approach consists in first aggregating the data using sliding windows, to obtain a sequence of usual static graphs. But this approach has many drawbacks, from the arbitrary choice of the duration of the time window, to the loss of temporal details inside the aggregated period, or the artifacts introduced by abrupt change at their arbitrary boundaries. Thus a recent trend of research (notably, stemming from \cite{latapy2018stream}) consists in designing concepts and methods able to work directly on those objects ---called here \textit{link streams}--- without having to use aggregation periods.

\subsubsection{Formal definition}
\begin{figure}
    \centering
    \includegraphics[width=1\textwidth]{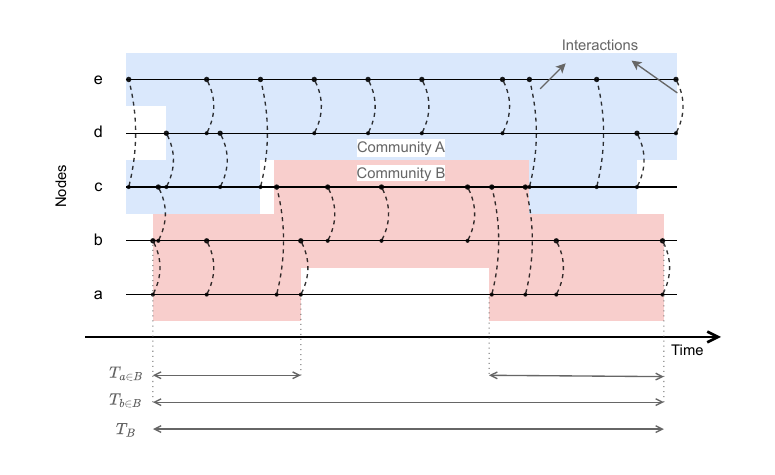}
\caption{Illustration of a link stream with a community structure depicted in blue and red. The figure includes representations of temporal community notations. The total interactions between nodes $b$ and $c$ are quantified as $L_{bc} = 4$. Within community $B$, these interactions total $L_{bc \in B}=3$. Additionally, the degree of node $a$ is represented as $k_a=8$.}
\label{fig:linkstream_example}
\end{figure}

\begin{definition}
A \textbf{link stream} $\mathcal{L}$ is defined by a triplet $(T, V, E)$ where $T \subset \mathbb{R}$ is a time interval, $V$ a finite set of $N \in \mathbb{N}$ nodes, and $E = \{(uv, t) \in V^2 \times T\}$ a finite set of interactions.
\end{definition}
In the following we focus on the case of simple dynamic graphs where interactions are instantaneous, undirected, and unweighted. Although this does not affect our contribution, $T$ is considered to be discrete.
To introduce the notations used in the following (see also Fig. \ref{fig:linkstream_example}):
\begin{itemize}
    \item Let $uv_t = 1$ if $(uv, t) \in E$, else $0$.              
    \item For a subset of time $T' \subset T$, define $L_{uv, T'} = \sum_{t \in T'} uv_t$ as the number of interactions between nodes $u$ and $v$ over $T'$.
    \item Let $L_{uv} = L_{uv, T}$ denote the total number of interactions between nodes $u$ and $v$.
    \item Similar to static graphs, let $k_u = \sum_{v \in V} L_{uv}$ denote the degree of a node $u$ and $m = \sum_{u \in V} k_u / 2 = |E|$ denote the total number of interactions in the link stream.
\end{itemize}

\subsection{Temporal Communities}

As there are multiple ways to define a temporal network, there are also multiple ways to define temporal communities. In static networks, communities are usually defined as complete partitions, i.e., a a set of set of nodes, such as each node belongs to exactly one community. 
In this article, we retain the principle of non-overlapping communities, but require 1) communities to be able to evolve with time, i.e., a community is a set of \textit{node-time} pairs $(u,t)$, 2)nodes to be able to belong to no community over some periods. Indeed nodes might have long periods of inactivity (e.g., nights in a fine-scale dataset, or even nodes that have left the systems, without us having this information, e.g., a phone number that is no longer attributed), and it would make little sense to try to include those inactive nodes in another partition, much as it would not make sense to include a node without edges in a static partition.

\begin{definition}\label{def:dyncom}
A dynamic community structure $\mathcal{C}$ over a link stream $L = (T, V, E)$ is a set of non-empty and mutually exclusive communities composed of sets of node-time pairs $\{(u_1, t_1), (u_1, t_2), ..., (u_2, t_3), (u_2, t_4), ...\}$.
\end{definition}

To introduce the notations used in the following:
\begin{itemize}
    \item $T_{u \in C} = \{t \in T \text{ s.t. } (u, t) \in C\}$ represents the times when a node $u$ belongs to community $C$.
    \item $T_{uv \in C} = T_{u \in C} \cap T_{v \in C}$ denotes the times nodes $u$ and $v$ simultaneously belong to community $C$.
    \item $L_{uv \in C} = L_{uv, T_{uv \in C}}$ represents the number of interactions between nodes $u$ and $v$ within community $C$.
    \item $T_{C} = \bigcup_{u \in V} T_{u \in C}$ denotes the existence time of community $C$.
    \item $C_u = \{C \in \mathcal{C} \mbox{ s. t. } T_{u \in C} \neq \emptyset\}$ denotes the set of communities visited by node $u$.
\end{itemize}

\section{Related work}
\label{sec:SOTA}
Detecting communities in temporal networks has been the focus of many previous research. However, most of these works are based on the so-called \textit{Instant-optimal} approach\cite{rossetti2018community}, i.e., first searching for static communities in a succession of snapshots, and then matching the partitions found using different strategies. The originality of our approach is that 1)We deal directly with link-stream data, without the need of aggregating into snapshots, and 2)We provide a quantitative definition of dynamic communities, that can be used as a quality function to evaluation a partition of a link stream. In this section, we will first discuss methods having only the first property, then those having only the second. Finally, we will see that the few methods at the intersection between the two suffer from important limitations.

\subsection{Community Detection on Stream Graphs, without objective function}
\subsubsection{Progressively Evolving Networks}

A few methods in the literature have proposed to get rid of snapshots by considering network evolution at the finest temporal granularity, updating communities at each network modification, such as node/edge addition/removal, e.g., \cite{cazabet2010detection,rossetti2017tiles,boudebza2018olcpm}. However, these methods are working on a particular type of temporal graphs, Progressively Evolving Graphs\cite{cazabet2020evaluating}, in which a well-behaved network exists at each point in time. This network is simply updated by adding or removing elements, but remains observable at any point in time, in a form on which static methods can be applied on it. This type of network corresponds for instance to \textit{relational data} such as friendships in a social media, or physical connections between routers in the internet. These methods are thus unable to deal with link streams, representing interactional data such as email, phone call, face-to-face interactions or messages in social media, in which the network exists only when considering interactions occurring at different times.

\subsubsection{Link Streams} 

A few methods have been proposed to search for persistent groups of nodes in Link Streams. In \cite{viard2016computing}, the authors propose to discover $\Delta$ cliques, i.e., groups of nodes that interact with each other at least once over a period $\Delta$. In \cite{boudebza2019detecting}, the authors search for groups of nodes that represent a consistently \textit{good} group ---in terms of \textit{conductance}--- over a sufficiently large period of time.

\subsection{Community Detection on Snapshot Sequences, optimizing a global objective function} 
Modern static methods for community detection mostly rely on a quantitative definition of what good communities are, expressed as an objective function, e.g., Modularity\cite{newman2004finding}, information compression on random walks\cite{rosvall2009map} or SBM inference \cite{peixoto2023descriptive}. Once this objective is provided, one can search for the partition optimizing it, usually through greedy heuristics, given the complexity of the problem. Methods in the previous section instead rely on ad-hoc rules, or the discovery of predefined patterns, such as cliques. The limit of these approaches is that they are not able to compare two partitions: they are only able to find 1)the only valid partition (pattern mining), or 2)a single partition as the result of a process.

Two approaches have proposed to adapt Modularity for temporal networks, i.e., having a single Modularity score for a temporal partition, thus explicitly incorporating the smoothing issue into their frameworks: the Average Modularity (A-Modularity) \cite{Aynaud2011MultiStepCD} and the Multislice Modularity (MS-Modularity) \cite{Mucha_2010}.

\subsubsection{Average Modularity} is defined only on a single, stationary partition of the temporal network, i.e., nodes are not allowed to join or leave communities along time. It is simply defined as the average modularity of the partition over all snapshots representing the graph.

%
\subsubsection{Multislice Modularity} \label{sec:mucha}is a more expressive adaptation of modularity, allowing nodes to change affiliation from one snapshot to the next. Its principle is to create a single multislice graph from a sequence of snapshots by adding artificial edges, known as interslice coupling, between successive temporal instances of each node. The authors introduce their formula for Multislice Modularity from the perspective of the flow stability of communities under a Laplacian dynamic \cite{delvenne2009stability}.
Noting $A_{ijs}$ the adjacency of nodes $i$ and $j$ at slice $s$, $k_{is}$ the degree of node $i$ at slice $s$, $m_s$ the number of edges in slice $s$, $\omega_{isr}$ the interslice weight between instances of slices $s$ and $r$ of node $i$, and $2\mu = \sum_s 2m_s + \sum_i \sum_{rs} \omega_{irs}$, MS-modularity is:
\begin{eqnarray}\label{eq:mod_mm}
    Q = \frac{1}{2\mu} \sum_{C \in \mathcal{C}} \sum_{i,j \in V^2} \sum_{r,s \in S^2}  \left[ \left( A_{ijs} -\frac{k_{is} k_{js}}{2m_s}\right) \delta_{sr} + \omega_{irs} \delta_{ij}\right]\delta_{ir \in C} \delta_{js \in C}
\end{eqnarray} 

MS-Modularity represents a convincing solution for temporal graphs provided as sequences of snapshots. However, as we will discuss when proposing our approach, it suffers from two limitations: 1)It works only on Progressively Evolving Graphs, and not on Link Streams, thus requiring to use snapshot aggregation as a preprocessing step, and 2)The results are highly dependent on the timescale chosen for aggregating into snapshots.

Beyond modularity, a few methods have been proposed to adapt the SBM inference approach to dynamic settings, such as \cite{yang2011detecting}. However, those methods are usually very costly to optimize, and cannot be adapted to link streams. A similar remark can be done for methods based on tensor decomposition, notably using the NMF \cite{ma2017evolutionary}. Their complexity becomes quickly intractable as the number of steps considered increases.

\subsection{Link streams and Objective Functions}

To the best of our knowledge, no version of the Modularity defined on link streams exists in the literature. At least one work has been conducted on SBM approaches\cite{matias2018semiparametric}, by considering that the probability of observing interactions between groups is modeled by a function of time. However, the authors show that to ensure identifiability, the partition must be stable, i.e., nodes cannot change communities. Another related work has been introduced in \cite{bovet2022flow}, based on the principle of flow stability. However, the method requires to choose a resolution parameter, and provide only two sets of partitions, the initial and final ones. 

Contrary to these methods, in this article we introduce L-Modularity, an adaptation of Modularity to link streams, allowing to assign a score for any partition of nodes of a link stream, including nodes changing communities, or even having no affiliation during some periods.

\section{Longitudinal modularity}\label{sec:longmod}
%


Modularity \cite{newman2004finding} is one of the most widely used methods for analyzing community structures in networks. Despite its known limitations, it is widely used for its intuitive definition, straightforward quality function, and ease of optimization. This simplicity makes it the ideal candidate for a first approach towards a quantitative definition of communities in link-streams.

Modularity of a community structure $\mathcal{C}$ over a network is based on the comparison of two terms: the observed number $\mathbb{L}_{\mathcal{C}}$ of edges within communities and their expected numbers $\mathbb{E}_{\mathcal{C}}$ based on a random null model. We can therefore express Modularity in a generic form as:
\begin{eqnarray}\label{eq:mod_stat1}
    Q = \mathbb{L}_{\mathcal{C}} - \mathbb{E}_{\mathcal{C}}
\end{eqnarray} 
The higher the Modularity, the more exceptional is the density inside communities, given the reference null model. Modularity was first introduced using the configuration model as reference \cite{newman2004finding}, which rewires edges while preserving the degrees. Denoting $A$ the adjacency matrix of an unweighted and undirected network, $k_i$ the degree of node $i$, $2m = \sum_{i} k_i$ twice the number of edges in the network, and $\mathcal{C}$ a community structure over it, the Modularity according to the configuration model is:
\begin{eqnarray}\label{eq:mod_stat2}
    Q = \frac{1}{2m} \sum_{C \in \mathcal{C}} \sum_{i, j \in V^2} \left[ A_{ij} - \frac{k_i k_j}{2m}\right] \delta_{i \in C} \delta_{j \in C}
\end{eqnarray} 

The selection of an appropriate null model is essential for defining modularity. Several null models have been proposed for static networks \cite{brissette2023effects,cazabet2017enhancing}, and significant efforts have been made to adapt Modularity for different types of networks. This includes weighted and directed networks \cite{Arenas_2007}, bipartite networks \cite{Barber_2007}, multi-layer networks \cite{paul2020nullmodelscommunitydetection}, and hypergraphs \cite{poda2024comparison}. As mentioned in section \ref{sec:mucha}, Modularity has also been adapted for multislice and temporal networks, in the context of sequence of snapshots\cite{Mucha_2010}. In the remaining of this section, we will show how we can formalize a generic form of temporal Modularity, as was done in Equation \ref{eq:mod_stat1} for the static one. We will then propose an instanciation of this generic formula for link streams, L-Modularity. 


\subsection{Abstact temporal modularity}


Multislice Modularity (later, MS-Modularity) was originally expressed (Eq. \ref{eq:mod_mm}) in a way which is specific to 1)a representation of dynamic graphs ---sequences of snapshots 2)a specific null model ---preserving degrees in each snapshots--- and 3)a particular way to ensure stable communities ---inter-snapshot edges with tunable weights.

We observe that, as we have done with static Modularity in \ref{eq:mod_stat1}, the general principle of a dynamic Modularity can be expressed in an abstract way as: the observed number of edges within communities $\mathbb{L}_{\mathcal{C}}$, minus their expected number $\mathbb{E}_{\mathcal{C}}$ based on a null model, plus a smoothness term $\mathbb{S}_{\mathcal{C}}$, penalizing nodes changing communities:
\begin{eqnarray}\label{eq:mod_mm2}
   Q = \mathbb{L}_{\mathcal{C}} - \mathbb{E}_{\mathcal{C}} - \mathbb{S}_{\mathcal{C}}
\end{eqnarray} 

In MS-Modularity(Eq. \ref{eq:mod_mm}), $\mathbb{L}_{\mathcal{C}}$ comes from the sum over edges present inside communities in each snapshot ($A_{ijs}$), the expected edges correspond to the expected number of edges in each snapshot given their degrees in that snapshot ($\frac{k_{is} k_{js}}{2m_s}$), and the smoothness term comes from inter-snapshot edges not being inside communities ($\omega_{irs}$).

We observe however that this particular instanciation of the abstract temporal Modularity of Eq. \ref{eq:mod_mm2} cannot work for link streams: since a valid graph does not exist at any given time $t$, it is not relevant to use the instantaneous degree, nor to represent a smoothness term as inter-temporal edges. Only $\mathbb{L}_{\mathcal{C}}$, i.e., the observed number of edges inside communities, remain a valid notion in link streams. As a consequence, we will in the following introduce instantiations of the expected number of edges $\mathbb{E}_{\mathcal{C}}$ and of the smoothness term $\mathbb{S}_{\mathcal{C}}$ tailored for link streams.


\subsection{Internal Edges Count $\mathbb{L}_{\mathcal{C}}$}

Given our defintions of link streams and communities (Sec. \ref{sec:definitions}), counting the number of edges inside communities is as simple as in static networks: an interaction $uv_t$ belongs to a community $C$ if both $u$ and $v$ belongs to $C$ at time $t$. Since we defined $L_{uv\in C}$ the number of interactions between $u$ and $v$ in community $C$, the total number of internal interactions can be computed as:
\begin{eqnarray}\label{eq:LC}
\mathbb{L}_{\mathcal{C}}=\sum_{u, v \in V^2}  \sum_{C \in \mathcal{C}} L_{uv \in C}
\end{eqnarray}

\subsection{Link Expectation Null Model}\label{sec:longitudinal_expectation}
%

In static Modularity, the most referenced null model is the configuration model, which randomizes the edges while preserving the degree sequence of the nodes \cite{newman2004finding}. With the notations of Eq. \ref{eq:mod_stat2}, the expectation of the presence of an edge between nodes $i$ and $j$ in a community $C \in \mathcal{C}$, according to this null model is given by 
\begin{eqnarray}\label{eq:statm_exp}
     \mathbb{E}\left[A_{ij}\right] = \dfrac{k_i k_j}{2m} \delta_{i \in C} \delta_{j \in C}
\end{eqnarray}
In other words, two nodes with high degrees are more likely to have a link between them. Conversely, observing a link between two nodes with low degrees is unexpected and may indicate the presence of a hidden community structure.

The expectation term of MS-Modularity in the sense of the abstract temporal modularity (Eq. \ref{eq:mod_mm2}) is a function of the degrees of the nodes in each snapshot. With the notations of Eq. \ref{eq:mod_mm}, the expected number of links between nodes $i$ and $j$ in a community $C$ over a set of snapshots $S$ is
\begin{eqnarray}\label{eq:mm_exp}
     \mathbb{E}\left[\sum_{s \in S} A_{ijs}\right] \propto \sum_{s \in S} \frac{k_{is} k_{js}}{2m_s} \delta_{is \in C} \delta_{js \in C}
\end{eqnarray}
In other words, the expected number of links is proportional to the sum of the expected numbers of links in each snapshot, with each snapshot's expectations determined by its specific configuration model. This approach is referred to as the SnapS null model in Gauvin et al. (2022) \cite{Gauvin_2022}.

In the context of link streams, we consider the null model denoted as $p[k, E]$ in Gauvin et al. (2022) \cite{Gauvin_2022}. This null model preserves the total degree of each node, but not the temporal sequence of interactions. 
For convenience, we name this null model the \textit{longitudinal configuration model}, as it directly generalizes the configuration model used for static networks. Note that this null model can be used both for link streams or snapshots. 

Compared with the null model used in MS-Modularity, it seems to better take into account the temporal nature of the data: if we imagine a temporal network in which a pair of nodes interact actively over short periods, while staying inactive during others, then using the Longitudinal Configuration Model as reference, we will naturally consider these periods of activity as exceptional. Instead, using MS-Modularity null model, the activity in each period is compared only with other nodes activity in that same period, and not with the activity of the same nodes in other periods.

The modularity expectation terms are typically derived directly from the null model, corresponding to the stationary probability of two random walkers arriving at each pair of nodes based on the transition probability given by the null model \cite{delvenne2009stability}. If we choose this approach, we must define transition probabilities between time instances for each node, thereby introducing artificial edges, which is precisely what we seek to avoid. 
We propose an alternative method, termed the \textit{longitudinal expectation} approach.
This approach constructs expectation terms based on the total duration of the link stream and multiplies them with temporal weights defined by node presence in communities. We present three versions of the expected number of interactions between nodes $u$ and $v$ in a community $C$.

\subsubsection{The co-membership expectation ($\mathbb{E}_{CM})$} (Eq. \ref{eq:longmonexpectation1}) 
represents the straightforward intuition that two nodes can only interact in time ranges in which they are both present.
\begin{eqnarray}\label{eq:longmonexpectation1}
     \mathbb{E}_{CM}\left[L_{uv \in C} \right] = \frac{k_{u} k_{v}}{2m} \frac{|T_{uv \in C}|}{|T|} 
\end{eqnarray}


\subsubsection{The joint membership expectation ($\mathbb{E}_{JM}$)} (Eq. \ref{eq:longmonexpectation2}), considers the overall structure of the community in a way that promotes stationary or nearly stationary communities, where nodes remain members for the entire duration of the community's existence. One could assume that, just as a "perfect" community in a static network is a clique disconnected from the rest of the graph, a "perfect" community in a link stream is a set of nodes with similar properties that remain unchanged—i.e., no node additions or removals—during its existence.
\begin{eqnarray}\label{eq:longmonexpectation2}
     \mathbb{E}_{JM}\left[L_{uv \in C} \right] = \dfrac{k_{u} k_{v}}{2m} \dfrac{|T_C|}{|T|} \mbox{  if } C \in C_u \cap C_v \mbox{, else } 0 
\end{eqnarray}

\subsubsection{The mean membership expectation ($\mathbb{E}_{MM}$)} (Eq. \ref{eq:longmonexpectation3}) expect the presence of edges to be proportional to the lifetimes of the two nodes within the community. It is calculated as the geometric mean of the two lifetimes.
\begin{eqnarray}\label{eq:longmonexpectation3}
     \mathbb{E}_{MM}\left[L_{uv \in C} \right] = \frac{k_{u} k_{v}}{2m} \frac{\sqrt{|T_{u \in C}| |T_{v \in C}|}}{|T|} 
\end{eqnarray}
Given that $|T_{uv \in C}| \leq \sqrt{|T_{u \in C}| |T_{v \in C}|} \leq |T_C|$, $\mathbb{E}_{MM}$ can be interpreted as a compromise between $\mathbb{E}_{CM}$ ---that may promotes overly erratic temporal communities--- and $\mathbb{E}_{JM}$ ---that promotes temporal communities with little or no evolution.

\vspace{0.5cm}
 We can note that, in the special case where two nodes belong to the community in the same time, $\mathbb{E}_{MM}$ is equivalent to $\mathbb{E}_{CM}$. Conversely, if one impose communities to stay unchanged during their time of existence, then $\mathbb{E}_{MM}= \mathbb{E}_{CM}=\mathbb{E}_{JM}$.
 Furthermore, if there is only one time step in the link stream, they are all equivalent to the expectation of the Modularity (Eq. \ref{eq:statm_exp}), which is in line with our objective of generalization.

\subsection{Smoothness term}\label{sec:disc_pen}

With no smoothness term for community discontinuities, modularity does not promote communities that are continuous over time (see Section \ref{sec:propdyncom}). 
We argue that the purpose of considering temporal community structures is to capture information on their lifecycle, which necessitates communities that are continuous over time.

MS-Modularity smoothness term is based on the creation of interslice couplings between successive temporal instances of each node, then sums the weights of these artificial edges that fall in-between communities, i.e., when a node changes affiliation between one snapshot and the next. As a consequence, the more nodes change community, the greater the term. Noting $\omega_{urs}$ the interslice weight of a node $u$ between snapshots $r$ and $s$, MS-Modularity smoothness term term is expressed as 
\begin{eqnarray}\label{eq:mm_timepen}
\mathbb{S}_{\mathcal{C}} & \propto \sum\limits_{C \in \mathcal{C}}\sum\limits_{u\in V} \sum\limits_{r,s \in \mathcal{S}^2} \omega_{urs}  \delta_{us \in C} \delta_{us \in C}
\end{eqnarray}
By definition, in temporal networks, $\omega_{urs} = 0$ if $|r-s| > 1$, i.e., when snapshots $r$ and $s$ are not successive. 

\begin{figure}
    \centering
    \includegraphics[width=0.75\textwidth]{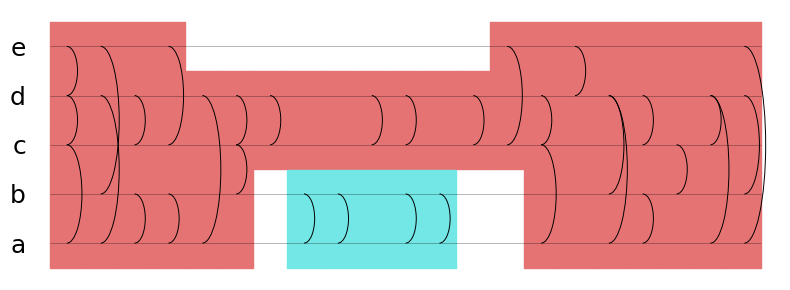}
\caption{\textbf{CSC Examples.} The figure is a 5 nodes link stream with 2 communities in red and blue. Nodes $a$ and $b$ leave the red community to form the two-nodes community in blue and then rejoin the red community again, so $\eta_a = \eta_b = 2$. Node $e$ temporarily leaves the red community, so $\eta_e = 1$. Since $c$ and $d$ never change communities, $\eta_c = \eta_d = 0$. Finally, $\rho=0.08$.}
\label{fig:acc_example}
\end{figure}

In our approach, it is essential to address the continuity of communities without relying on artificial temporal edges between time steps. To achieve this, we propose calculating the \textbf{Community Switch Count} (CSC), denoted as $\eta_u$, for each node $u$. The CSC represents the number of times a node transitions out of a community and subsequently joins another community. More precisely, $\eta_u$ is the number of communities visited by node $u$, minus one. This count also accounts for instances where a node revisits the same community multiple times. Fig. \ref{fig:acc_example} illustrates examples of the CSC. In the critical case where nodes change communities at each interaction, the following inequality holds: $0 \leq  \eta_u \leq k_u - 1$ for each node $u$, and therefore $0 \leq \sum_{u \in V} \eta_u \leq 2m - N$.

To maintain generality 
, we propose measuring the time discontinuity of a dynamic community structure with 
\begin{eqnarray}\label{eq:lm_timepen}
     \rho(L, \mathcal{C}) = \frac{1}{2m} \sum_{u \in V}\eta_{u}(\mathcal{C})
\end{eqnarray}

We propose using this measure as a smoothness term $\mathbb{S_{\mathcal{C}}}$ for L-Modularity. In the scenario where there is only one time step in the link stream, $\rho = 0$, effectively generalizing Modularity. For the edge case mentioned above, $\rho$ approaches $1$. If there are even more discontinuities, $\rho$ is not bounded. 
The smoothness term can be weighted according to specific requirements. A higher weight value results in greater penalization of discontinuities. For instance, with a weight of $2m$ and only one node changing community, L-Modularity will be less than $0$, thereby promoting stationary communities. Similarly, with a weight of $2$, L-Modularity will be less than $0$ if, on average, nodes change community every two interactions. The weighting of the smoothness term may be refined, and its impacts should be explored in future research.

%






\subsection{Extension to Multigraphs and Weighted graphs}
Static Modularity naturally extends to multigraphs, by considering that the node degrees $k_u$ correspond to the total number of edges, eventually repeated. Similarly, it naturally generalises to weighted graphs by using the node strengths (i.e., sum of weights of adjacent nodes) as values for $k_u$, and normalizing accordingly.

The same generalization can be done for L-Modularity. This generalization is particularly relevant in contexts in which temporal networks have been obtained by aggregating observations over periods of time, e.g., counting social interactions between people over a day, a week, or a month. In this situation, one can represent the strengths of these observations using multigraphs or weights. Note that when generalizing to any type of weighted graphs, in particular with weights lower than 1, it might be necessary to renormalize the smoothness term. 

We will show in section \ref{sec:independence} that, unlike MS-Modularity, L-Modularity provide coherent results in aggregated graphs compared with their non-aggregated versions (\textit{Independence to time-aggregation property}).


%
\section{Properties of dynamic communities}\label{sec:propdyncom}
%

In the previous section, we proposed an adaptation of Modularity to link streams. In this section, we propose several properties that seem desirable for a definition of dynamic community structures in temporal networks
and position L-Modularity relative to these properties. 


Although the exact definition of what are \textit{good} communities in static networks remains an open discussion, most authors agree on a general idea of groups \textit{more strongly connected internally than the rest of the network}. Modularity is a mathematical transcription of this general principle ---one among others \cite{HOLLAND1983109} \cite{vonluxburg2007tutorialspectralclustering} \cite{Rosvall_2009}\cite{peixoto2023descriptive}. One way to ensure that such a quantitative definition is compatible with the original intuition is to check that it respects some desired properties. For instance, one could say that Modularity in static networks respects the following properties:
\begin{itemize}
    \item The score does not favor \textit{trivial} solutions, such as having all nodes in the same community or each node in its own community.
    \item If there are two separated connected components in the graph, Modularity necessarily attributes a lower score to the partition merging them than to another one identical in other aspects, but keeping them independent.
    \item Communities are always denser than the overall network, as long as a solution is found with Modularity $Q>0$.
\end{itemize}
We define similar properties for quality functions for link stream partitions.

\begin{figure}[!t]
\centering
\begin{subfigure}[b]{0.9\textwidth}
         \centering
         \includegraphics[width=0.9\linewidth]{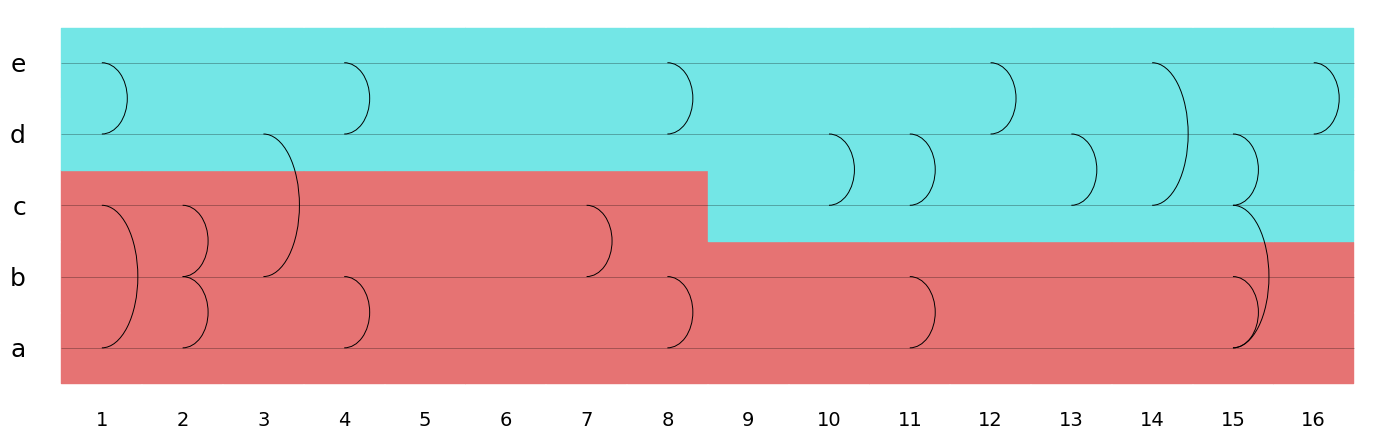}
         \caption{Link stream, time granularity $\Delta_t=1$}
     
\end{subfigure}
\begin{subfigure}[b]{0.9\textwidth}
\centering
\includegraphics[width=0.9\linewidth]{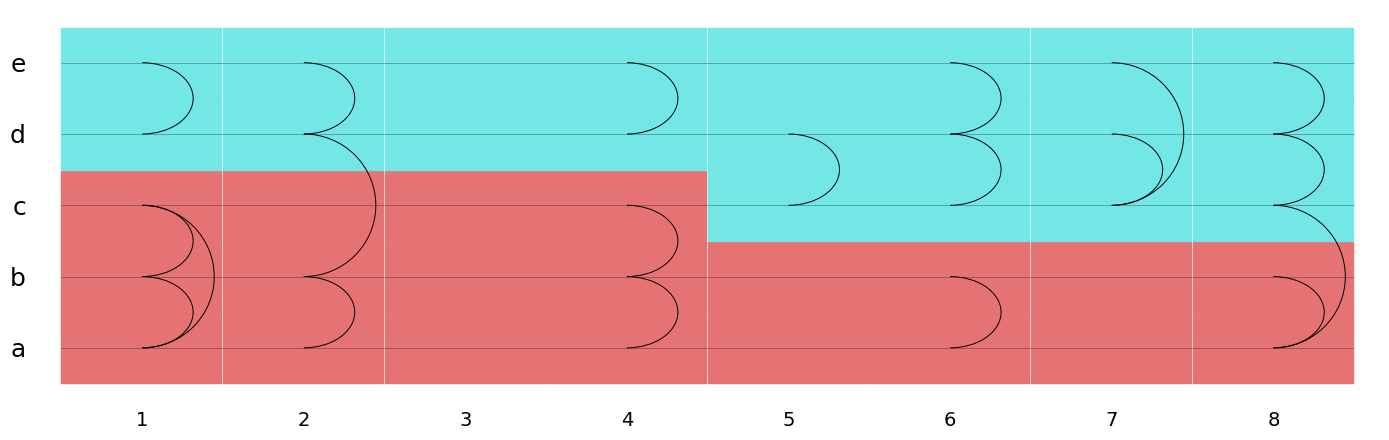}
\caption{Link stream, time granularity $\Delta_t=2$}
\end{subfigure}
 \begin{subfigure}[b]{0.9\textwidth}
 \centering
 \includegraphics[width=0.85\linewidth]{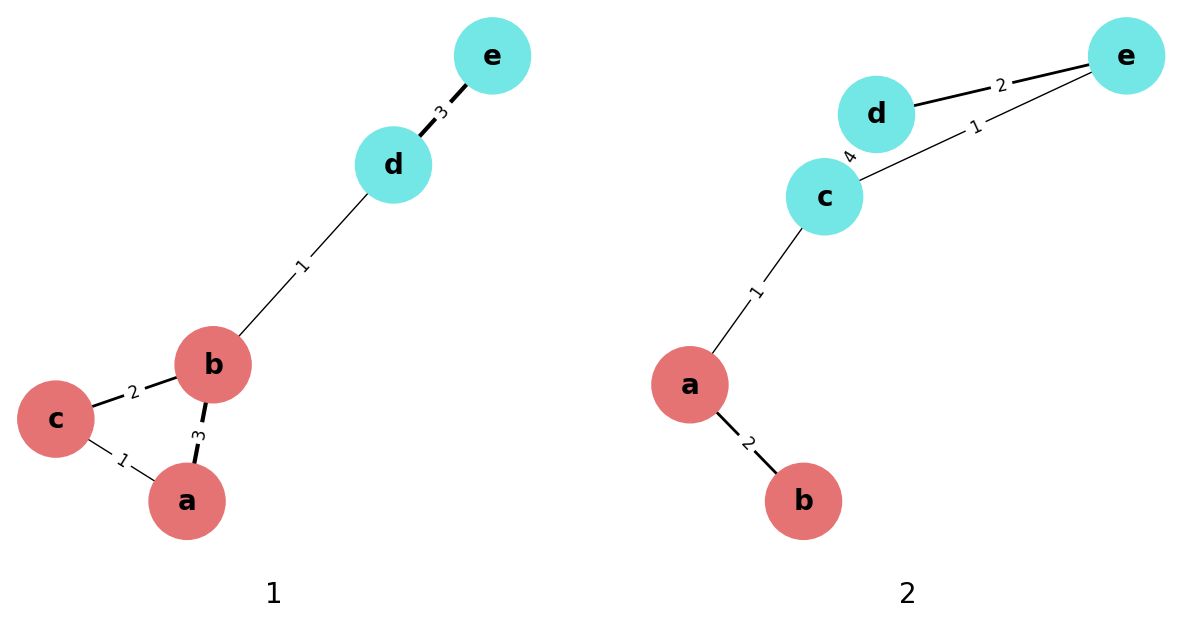}
         \caption{Snapshots, time granularity $\Delta_t=8$}
\end{subfigure}
\caption{Illustration of the independence of L-Modularity from the time granularity. All 3 examples are composed of the same interactions occurring at the same time, but aggregated at different time scales. The L-Modularity score is the same for the blue/red communities in all 3 cases, since 1)the number of interactions inside each community, 2)the relative duration of communities, 3)the global nodes degree, and 4)the number of affiliation discontinuity are all identical in all 3 cases.}
\label{fig:freq_indep}
\end{figure}

\subsection{Property 1: Independence to time-aggregation property}
\label{sec:independence}
A common challenge in the study of temporal networks is defining a suitable time window for aggregation or determining an appropriate time granularity prior to performing dynamic community detection \cite{nonalter} \cite{quantifyingtheeffets} \cite{Krings_2012} \cite{Scholtes_2016} \cite{caceres_2011}.

A notable feature of L-Modularity is that it yields the same score for a temporal node partition on different representations of the same link-stream obtained by different time-window aggregations without loss of information, as illustrated in Fig. \ref{fig:freq_indep}. This property can be expressed as follows:

\begin{definition}\label{def:time-aggregation}
A dynamic community quality function is said to be \textbf{independent to time-aggregation} if, for two representations $L$ and $L'$, such as $L'$ is obtained by aggregating $L$ into static graphs by time intervals ---multiple interactions over a period being representing by weights or multi-edges--- a partition defined on $L'$ yields the same score on $L$.
\end{definition}

A significant advantage of this property is that it makes L-Modularity a useful tool for comparing dynamic community structures on the same temporal network aggregated with different window sizes (see Section \ref{sec:experimentation}). In contrast, different MS-Modularity scores for community structures on the same temporal network with different aggregation window sizes are not comparable, as the multislice network changes each time.

\begin{figure}[!t]
\centering
\resizebox{0.925\textwidth}{!}{
\begin{subfigure}[b]{0.49\textwidth}
         \centering
         \includegraphics[width=\textwidth]{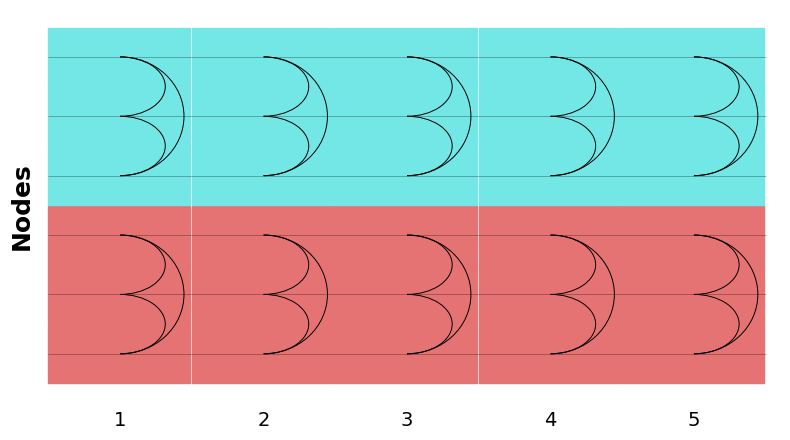}
         \caption{Continuous}
         \label{fig:y equals x}
     \end{subfigure}
\begin{subfigure}[b]{0.49\textwidth}
         \centering
         \includegraphics[width=\textwidth]{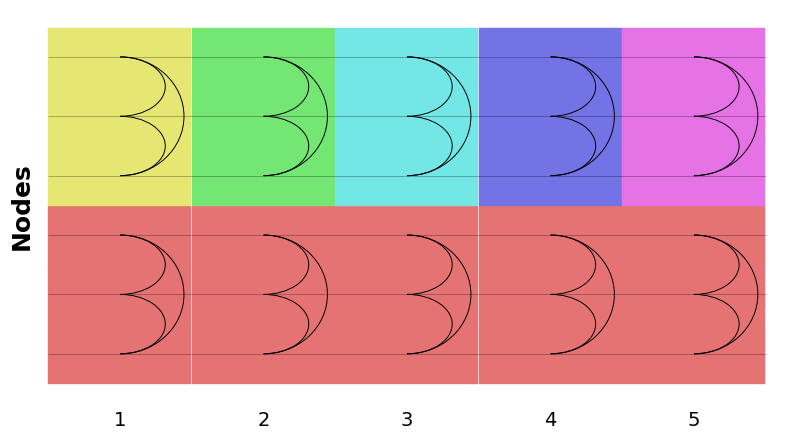}
         \caption{Non-continuous}
         \label{fig:y equals x}
     \end{subfigure}
     }
\caption{Illustration of the \textit{smoothness incentive} property. The same link stream is partitioned in different ways, each color representing a community. In Fig. (b), the three nodes on the top change communities, although there is not change in the network structure. A quality function respecting the smoothness incentive property must strictly favor partition (a) over (b)}
\label{fig:continuity_property}
\end{figure}

\subsection{Property 2: Smoothness Incentive}\label{sec:stability}

Addressing the smoothness of temporal communities is a well-known challenge in the field of dynamic community detection \cite{cazabet2020evaluating}. We propose that a quality function for temporal community structures must explicitly promote this smoothness, as illustrated in Fig. \ref{fig:continuity_property}. More formally: 

\begin{definition}\label{def:continuity_property}
Let's consider a temporal network $L$ studied over a period $T$, split into two-time intervals $T_1$ and $T_2$, and for which 1) the cumulated networks on both intervals are identical, 2)the partitions $\mathcal{C_1}$ on $T1$ and $\mathcal{C_2}$ on $T2$ in each interval are identical. A dynamic community quality function is said to have the \textbf{smoothness incentive} if, for a partition on $T$, it yields a strictly higher score to a partition in which each community $C_i\in \mathcal{C_1}$ is merged with its corresponding community $C_i\in \mathcal{C_2}$, than to a partition in which any community in $\mathcal{C_1}$ remains distinct from its equivalent one in $\mathcal{C_2}$
\end{definition}

\subsection{Property 3: Topochrone Disconnection Property}

In static networks, non-connected components indicate that two subsets of nodes have no links between them. Modularity assigns a higher score when non-connected components belong to different communities.
In temporal networks, \textit{disconnected topochrone components} (Def. \ref{def:disco_topo}) occur when two sub-link streams do not share any nodes nor time intervals of existence. We propose that a quality function for temporal community structures should assign a higher score when disconnected topochrone components belong to different communities, rather than the same one.

\begin{definition}\label{def:disco_topo}
Two sub link streams $L_1$ and $L_2$ are said to be two \textbf{disconnected topochrone components} of a link stream if $T_1 \cap T_2 = V_1 \cap V_2 = \emptyset$.
\end{definition}

\begin{definition}\label{def:disc_topo_comp}
A quality function for dynamic communities is said to respect the \textbf{topochrone disconnection} property if, given a partition in which a community contains two disconnected topochrone components, it gives a strictly higher score to the same partition in which those two topochrone components are split in two separate communities.
\end{definition}

\begin{figure}[!htb]
\centering
\resizebox{1\textwidth}{!}{
\begin{subfigure}[b]{0.49\textwidth}
         \centering
         \includegraphics[width=\textwidth]{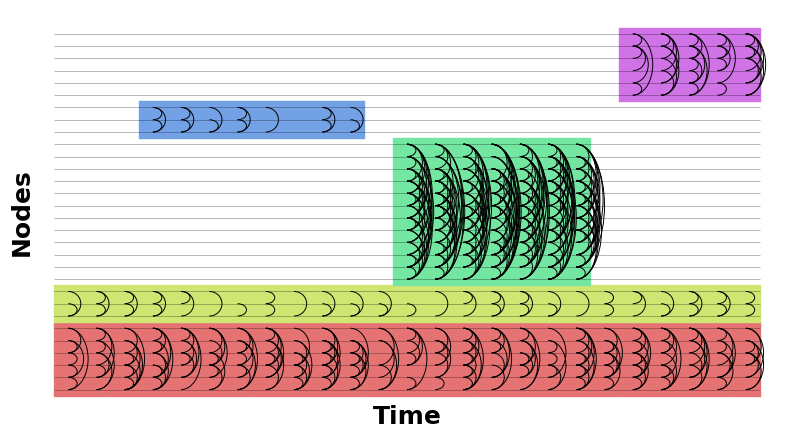}
         \caption{}
         \label{fig:topodis_property_a}
     \end{subfigure}
\begin{subfigure}[b]{0.49\textwidth}
         \centering
         \includegraphics[width=\textwidth]{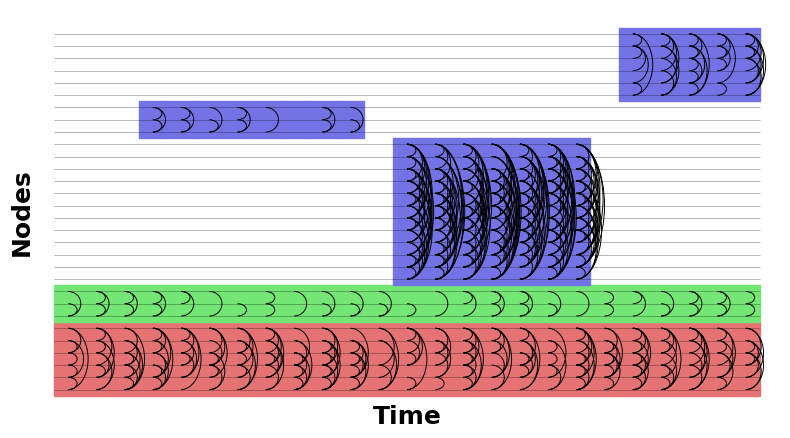}
         \caption{}
         \label{fig:topodic_property_b}
     \end{subfigure}
     }
\caption{\textbf{Topochrone disconnections.} A quality function respecting the Topochrone disconnection property should strictly favor the partition in Fig. a) over the one in b).}

\label{fig:topodis_property}
\end{figure}

For example, the three communities shown in figure \ref{fig:topodis_property_a} in blue, green, and purple are three disconnected topochrone components. A quality function respecting the topochrone disconnection property should strictly favor partition \ref{fig:topodis_property_a} over \ref{fig:topodic_property_b}.

\subsection{L-Modularity and temporal community properties}
As discussed in Section \ref{sec:longmod}, L-Modularity results from combinations of the inclusion or non-inclusion of the smoothness term and the choice of longitudinal expectation. We will see which combinations respect the properties.

\subsubsection{L-Modularity and Independence to time-aggregation}
L-Modularity is independent of time-aggregation, while MS-Modularity isn't.

This property naturally emerges from the elements used to compute L-Modularity, which depends only on the number of interactions inside communities, the time periods during which nodes belong to communities, and the total number of interactions per node.

On the contrary, MS-Modularity depends 1)on the number of aggregation steps, since a higher number of snapshots leads to a higher number of inter-slice edges, encoding the smoothness term and 2)on the local degree of nodes at each step, and thus on the simultaneity of interactions, which is modified by aggregating.

It is important to note that while L-Modularity yields the same value for the same partition of the same link stream provided at multiple temporal granularities, the optimum partition might be different since a finer granularity allows to express partitions that would not be possible at coarser ones.

\subsubsection{L-Modularity and Smoothness Incentive}

The smoothness incentive property is insured by the smoothness term $\rho$ (Eq. \ref{eq:lm_timepen}), that penalizes unnecessary changes in affiliation.

\subsubsection{L-Modularity and Topochrone Disconnection}
This property is insured by the expectation term of L-Modularity. Indeed, when using $\mathbb{E}_{MM}$ (or $\mathbb{E}_{JM}$), the number of edges expected inside a community grows with the time spent by nodes inside communities, even if this presence is not simultaneous. This corresponds to a vision of a perfect dynamic community as a set of nodes being homogeneously connected over a period. The topochrone disconnection goes against this objective.
Note that for MS-Modularity, the situation relative to Topochrone disconnection is unclear and depends on modeling choices. If a node inactive at time $t$ is removed completely from the network, then MS-Modularity does not fulfill the topochrone disconnection property, since no interslice edge will be added by merging the communities in a single one. If inactive nodes remain in the graph, then it respects this property, but will cause other difficulties, such as artificially rewarding the stability of communities without any observed internal edges.

\subsection{Discussion on temporal communities properties}

L-Modularity values for different term combinations have been calculated for all community structures shown in Figures \ref{fig:continuity_property} and \ref{fig:topodis_property}. 
The results are summarized in table \ref{table:properties_timepen}, which illustrates which properties are promoted by the three expectations and the presence or absence of the smoothness term. 
The results indicate the necessity of a smoothness term to promote smoothness incentives. Additionally, it is noted that the co-membership expectation (Eq. \ref{eq:longmonexpectation1}) does not promote robustness to topochrone disconnections.
Therefore, we propose using either the joint membership expectation (Eq. \ref{eq:longmonexpectation2}) or the mean membership expectation (Eq. \ref{eq:longmonexpectation3}), along with the smoothness term (Eq. \ref{eq:lm_timepen}).

\begin{table}
    \begin{center}
\[
\begin{array}{| c || c | c | c | }
    \hline
    & \; \textbf{Independance to} \;& \;\textbf{Smoothness} \; & \textbf{Robust to topo-} \\
    & \;\textbf{time aggregation} \;& \textbf{Incentive} & \; \textbf{chrone disconnections} \; \\
    \hline\hline
    
    \mathbb{E}_{CM} & \large\textcolor{green!50!black}{\checkmark}& \large\textcolor{red}{\times}& \large\textcolor{red}{\times} \\
    
    \hline
    
    \hspace{0.2cm} \mathbb{E}_{CM}+ \rho \hspace{0.2cm}& \large\textcolor{green!50!black}{\checkmark}& \large\textcolor{green!50!black}{\checkmark} & \large\textcolor{red}{\times} \\
    
    \hline
    
    \mathbb{E}_{JM} & \large\textcolor{green!50!black}{\checkmark}& \large\textcolor{red}{\times} & \large\textcolor{green!50!black}{\checkmark}  \\
    \hline

    \mathbb{E}_{JM}+ \rho & \large\textcolor{green!50!black}{\checkmark}& \large\textcolor{green!50!black}{\checkmark} & \large\textcolor{green!50!black}{\checkmark} \\
\hline
    
    \mathbb{E}_{MM} & \large\textcolor{green!50!black}{\checkmark}& \large\textcolor{red}{\times} & \large\textcolor{green!50!black}{\checkmark} \\
    \hline

    \mathbb{E}_{MM}+ \rho & \large\textcolor{green!50!black}{\checkmark}& \large\textcolor{green!50!black}{\checkmark} & \large\textcolor{green!50!black}{\checkmark} \\
    \hline
    
    \hline
\end{array}
\]
    \end{center}
    \caption{Properties promoted by the different terms introduced in Section \ref{sec:longmod}: the expectations $\mathbb{E}_{CM}$,  $\mathbb{E}_{JM}$,  $\mathbb{E}_{MM}$  and the smoothness term $\rho$. 
    }
        \label{table:properties_timepen}
\end{table}

\section{Experimentation}\label{sec:experimentation}

In this section, we propose to evaluate L-Modularity using two different longitudinal expectations: the joint membership expectation (denoted as $Q_{\mathcal{C}, JM}$, based on Eq. \ref{eq:longmonexpectation2}) and the mean membership (denoted as $Q_{\mathcal{C}, MM}$, based on Eq. \ref{eq:longmonexpectation3}), both with a smoothness weight of $2$. Specifically, we use the SocioPatterns high school dataset \cite{Mastrandrea_2015}, which captures the contacts and friendship relations between students at a high school in Marseilles, France, during December 2013. The interactions are recorded as 20-second interval contacts among students from nine classes over five days. The dataset also includes information on the class membership of each student.

We apply two methods to uncover dynamic community structures within the dataset. The first method involves optimizing MS-Modularity, varying the interslice weight ratio. 
The second method employs a no-smoothing approach \cite{cazabet2020evaluating} (later, NS-Modularity) which first applies the Louvain algorithm \cite{Blondel_2008} on each snapshot and then merges successive communities based on the Jaccard Index of their node sets. We experimented the NS-Modularity method varying the minimum threshold on the Jaccard Index for two communities to merge. Ultimately, we compute the two versions of L-Modularity for each community structure obtained. 

\begin{figure}[!hbt]
\centering
\begin{subfigure}[b]{.49\textwidth}
         \centering
         \includegraphics[width=0.9\textwidth]{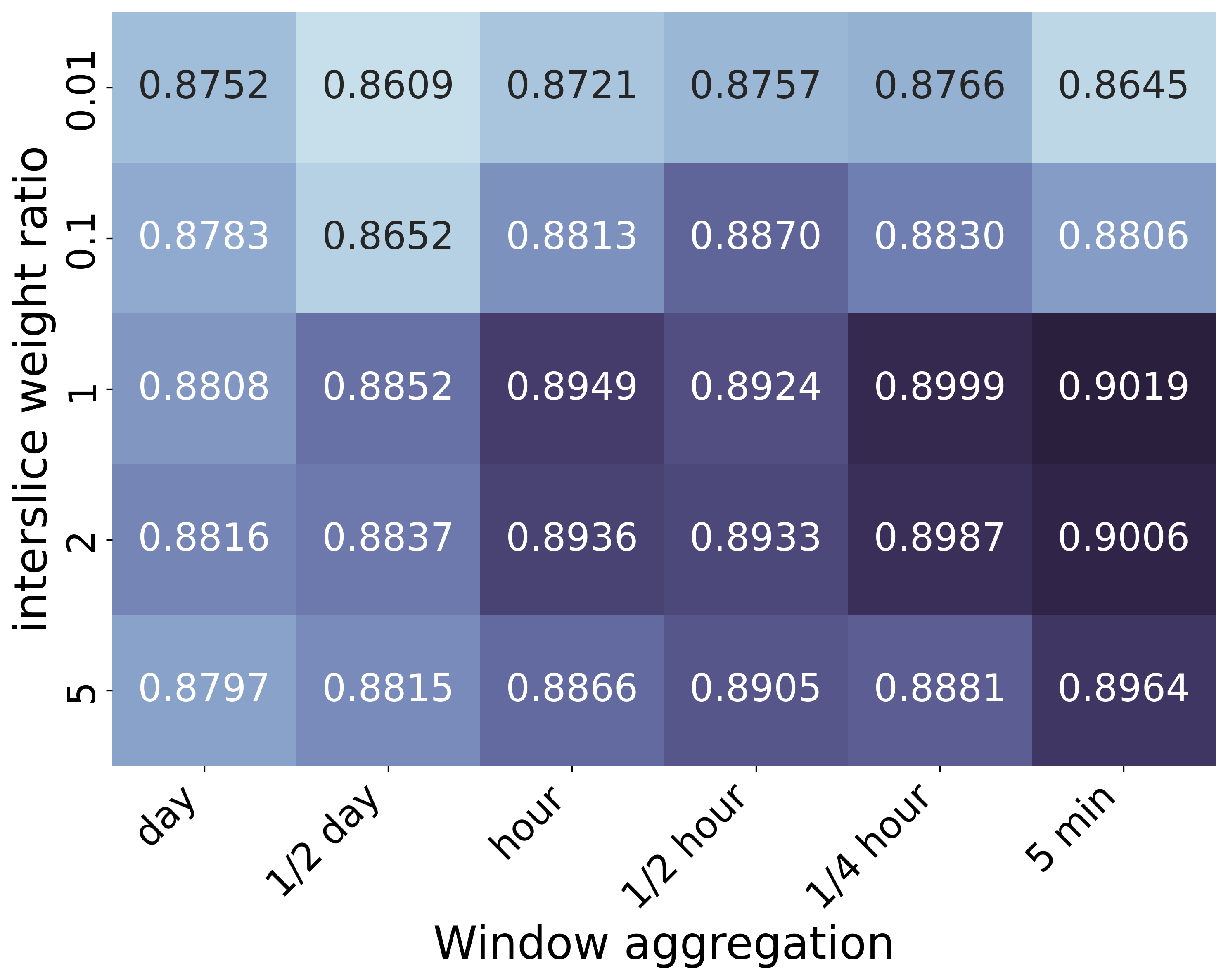}
         \caption{\fontsize{9}{11}\selectfont MS method, evaluated by $Q_{\mathcal{L}, MM}$}
    \label{fig:ms_mm_heatmap}
    \medskip
     \end{subfigure}
\begin{subfigure}[b]{.49\textwidth}
         \centering
         \includegraphics[width=0.9\textwidth]{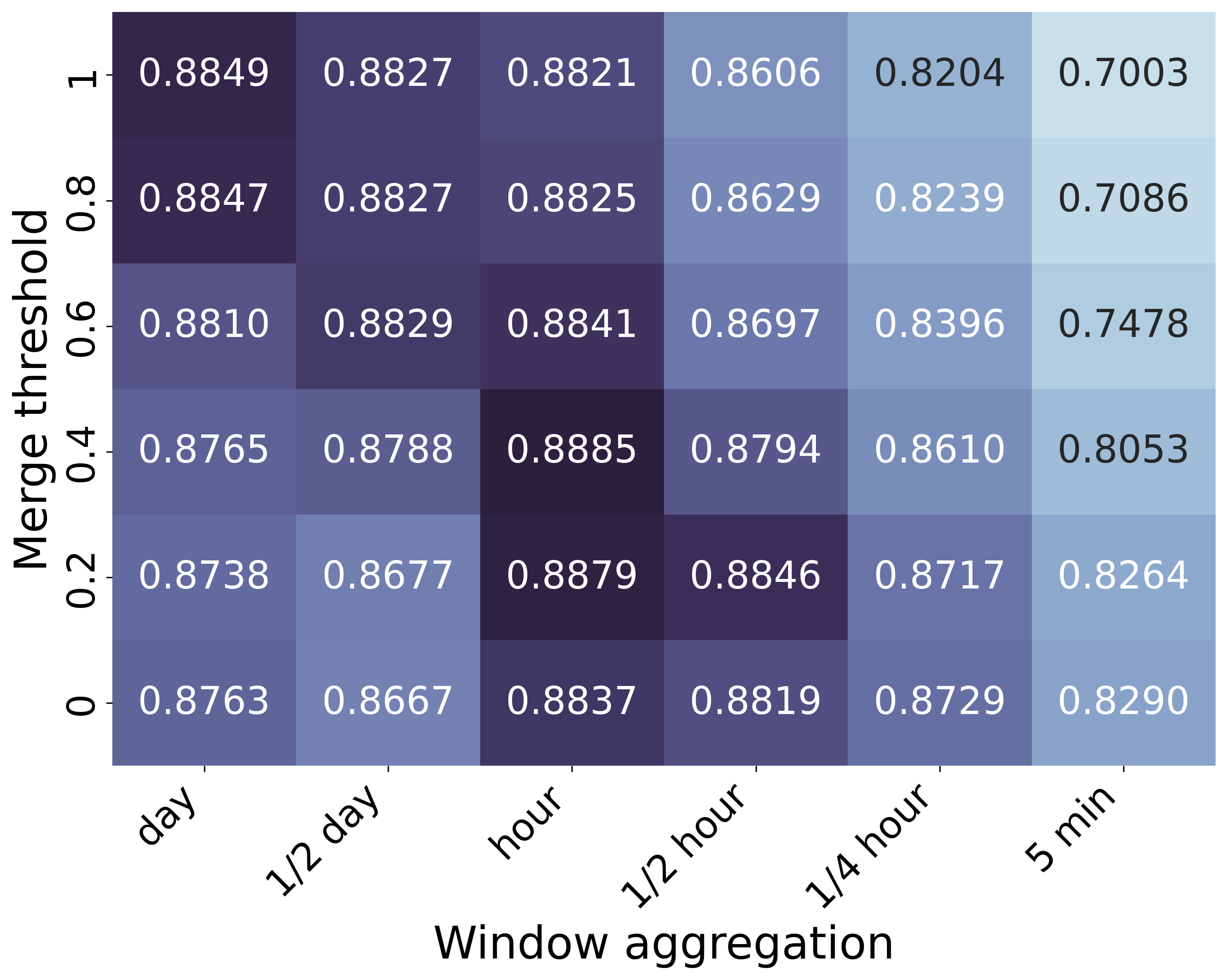}
         \caption{\fontsize{9}{11}\selectfont NS method, evaluated by $Q_{\mathcal{L}, MM}$}
         \label{fig:ns_mm_heatmap}
         \medskip
     \end{subfigure}
\begin{subfigure}[b]{.49\textwidth}
         \centering
         \includegraphics[width=0.9\textwidth]{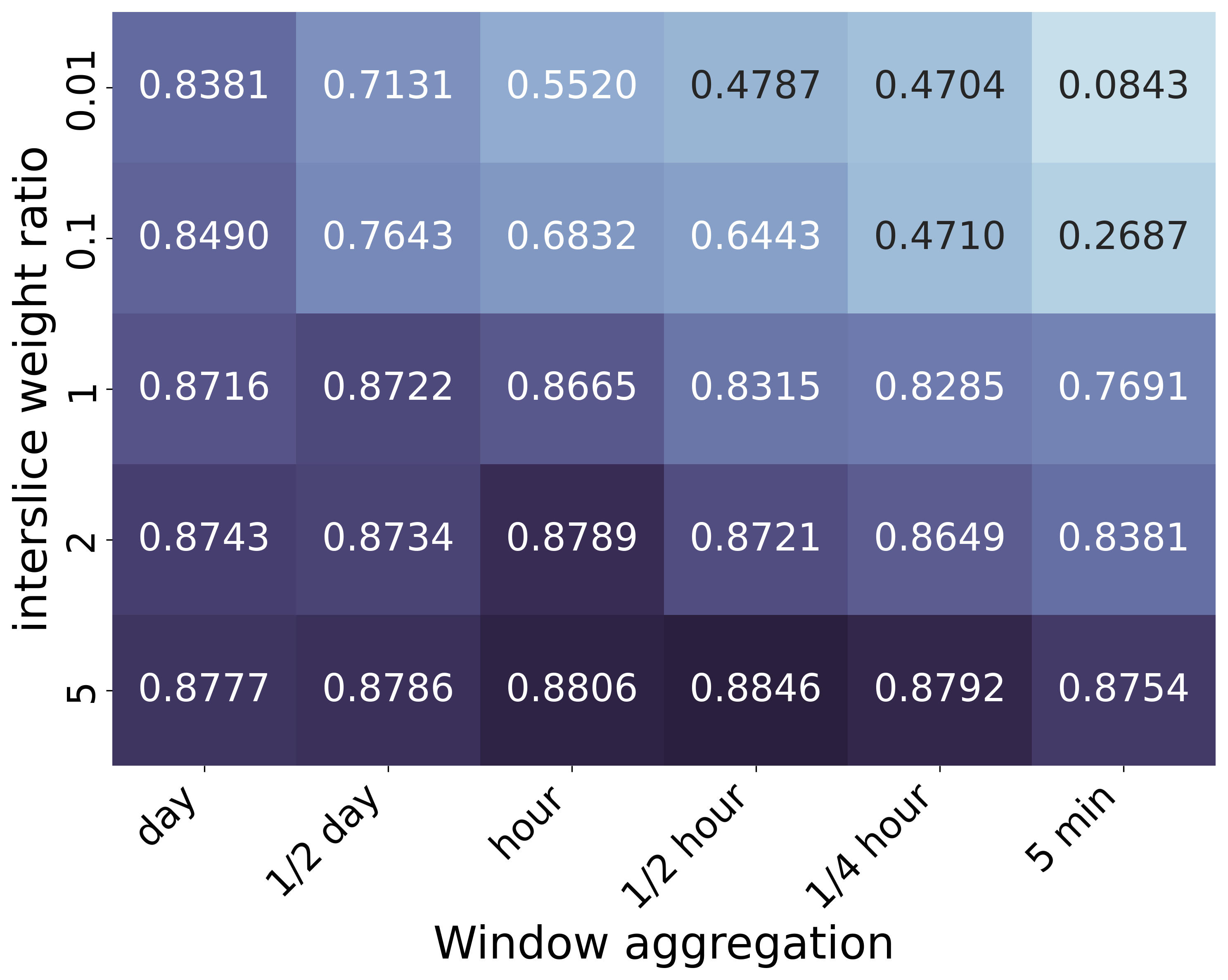}
         \caption{\fontsize{9}{11}\selectfont MS method, evaluated by $Q_{\mathcal{L}, JM}$}
    \label{fig:ms_jm_heatmap}
     \end{subfigure}
\begin{subfigure}[b]{.49\textwidth}
         \centering
         \includegraphics[width=0.9\textwidth]{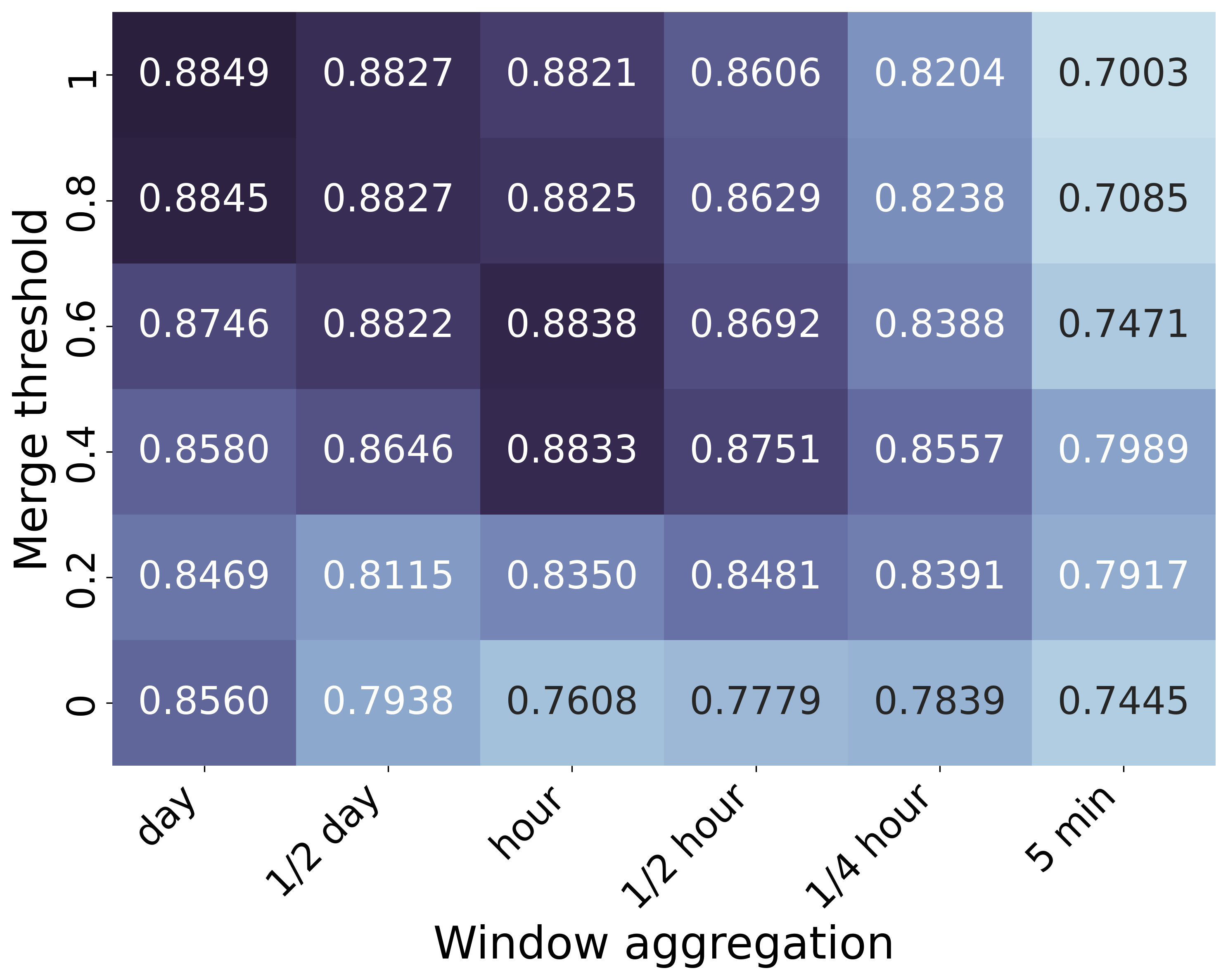}
         \caption{\fontsize{9}{11}\selectfont NS method, evaluated by $Q_{\mathcal{L}, JM}$}
    \label{fig:ns_jm_heatmap}
     \end{subfigure}
\caption{Heatmaps of L-Modularity scores of dynamic community structures revealed by MS-Modularity optimization (MS) and No-smoothing method (NS) with varying parameters. Evaluations were conducted using two versions of L-Modularity: one, $Q_{\mathcal{L}, MM}$, with the mean membership expectation and one, $Q_{\mathcal{L}, JM}$, with the joint membership expectation.}
\label{fig:sociopatternhighchool_heatmaps}
\end{figure}

\begin{figure}[!hbt]
\centering
\begin{subfigure}[b]{0.49\textwidth}
         \centering
         \includegraphics[width=0.9\textwidth]{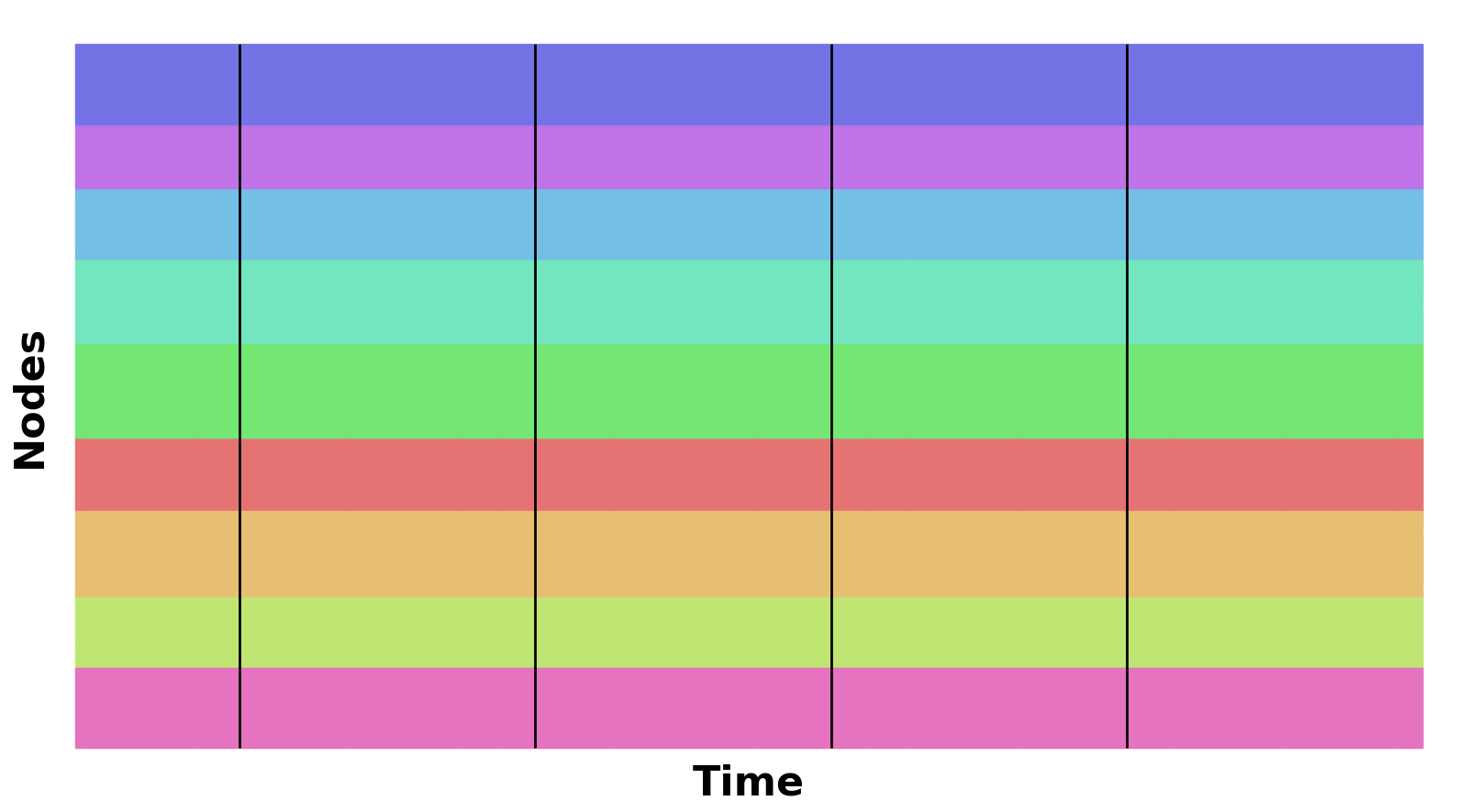}
         \captionsetup{width=0.95\textwidth}
         \caption{\fontsize{8}{9.5}\selectfont Students classes as communities. $Q_{\mathcal{L}, MM} = Q_{\mathcal{L}, JM} = 0.8718$.}
     \label{fig:classes}
     \medskip 
     \end{subfigure}
     \begin{subfigure}[b]{0.49\textwidth}
         \centering
         \includegraphics[width=0.9\textwidth]{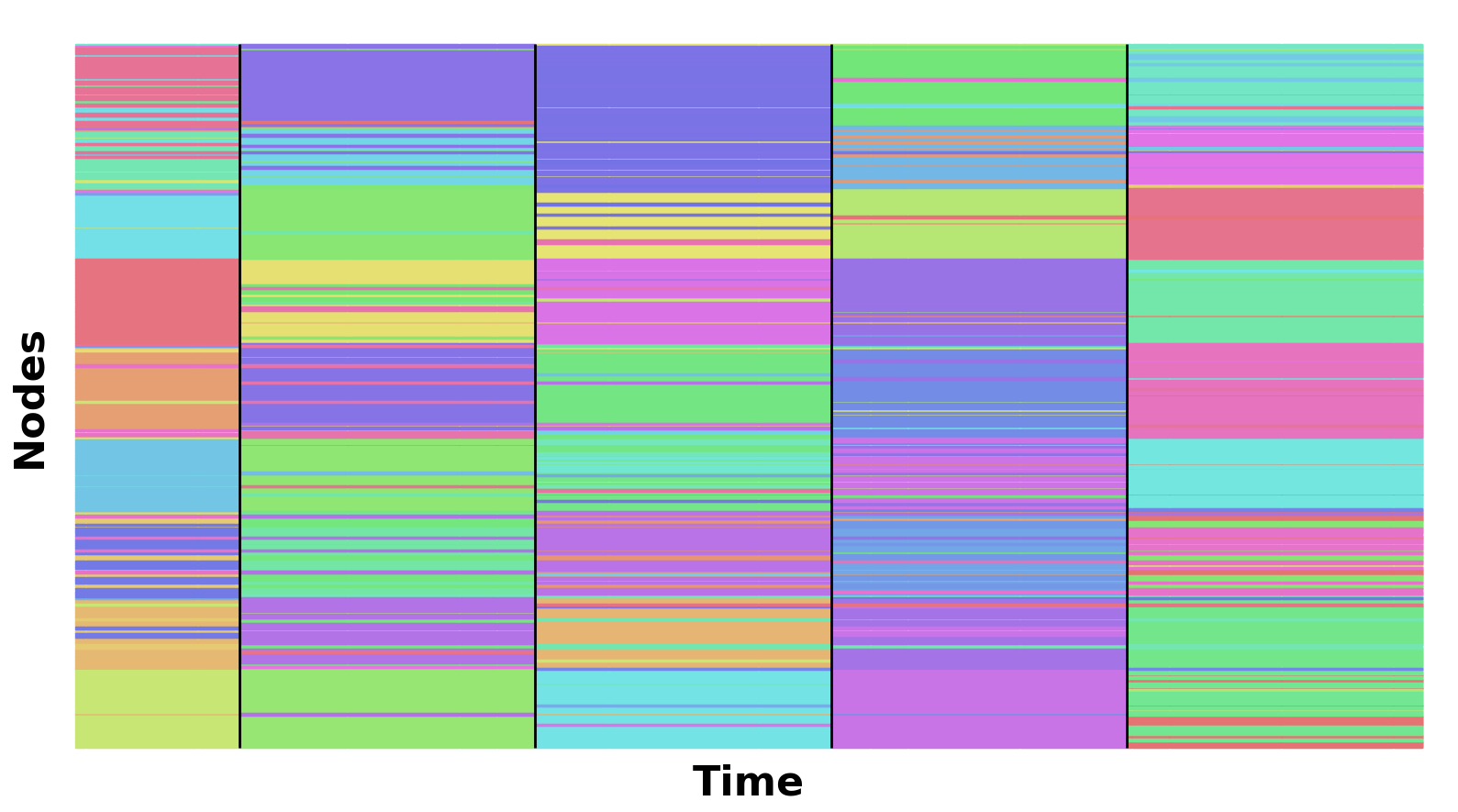}
         \captionsetup{width=0.95\textwidth}
         \caption{\fontsize{8}{9.5}\selectfont The optimal structure for the joint membership expectation with 157 communities: $Q_{\mathcal{L}, JM} = 0.8849$.}
    \label{fig:ms_mm_best}
     \end{subfigure}
    \begin{subfigure}[b]{0.49\textwidth}
         \centering
         \includegraphics[width=0.9\textwidth]{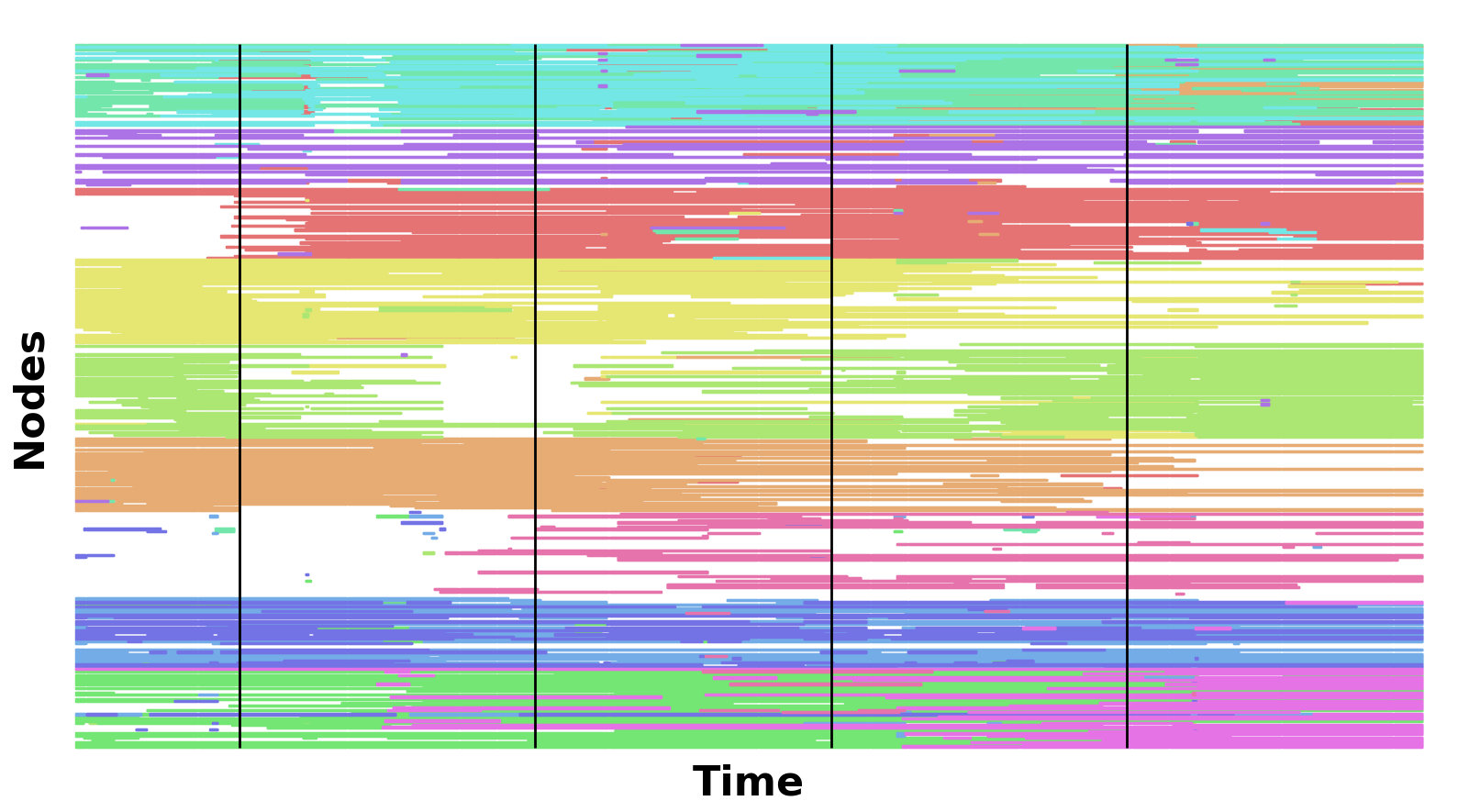}
         \captionsetup{width=0.985\textwidth}
         \caption{\fontsize{8}{9.5}\selectfont The optimal structure for the mean membership expectation: $Q_{\mathcal{L}, MM} = 0.9019$. For clarity, only the top 12 most represented communities out of the 48 identified are displayed.}
    \label{fig:ms_jm_best}
     \end{subfigure}
\begin{subfigure}[b]{0.49\textwidth}
         \centering
         \includegraphics[width=0.9\textwidth]{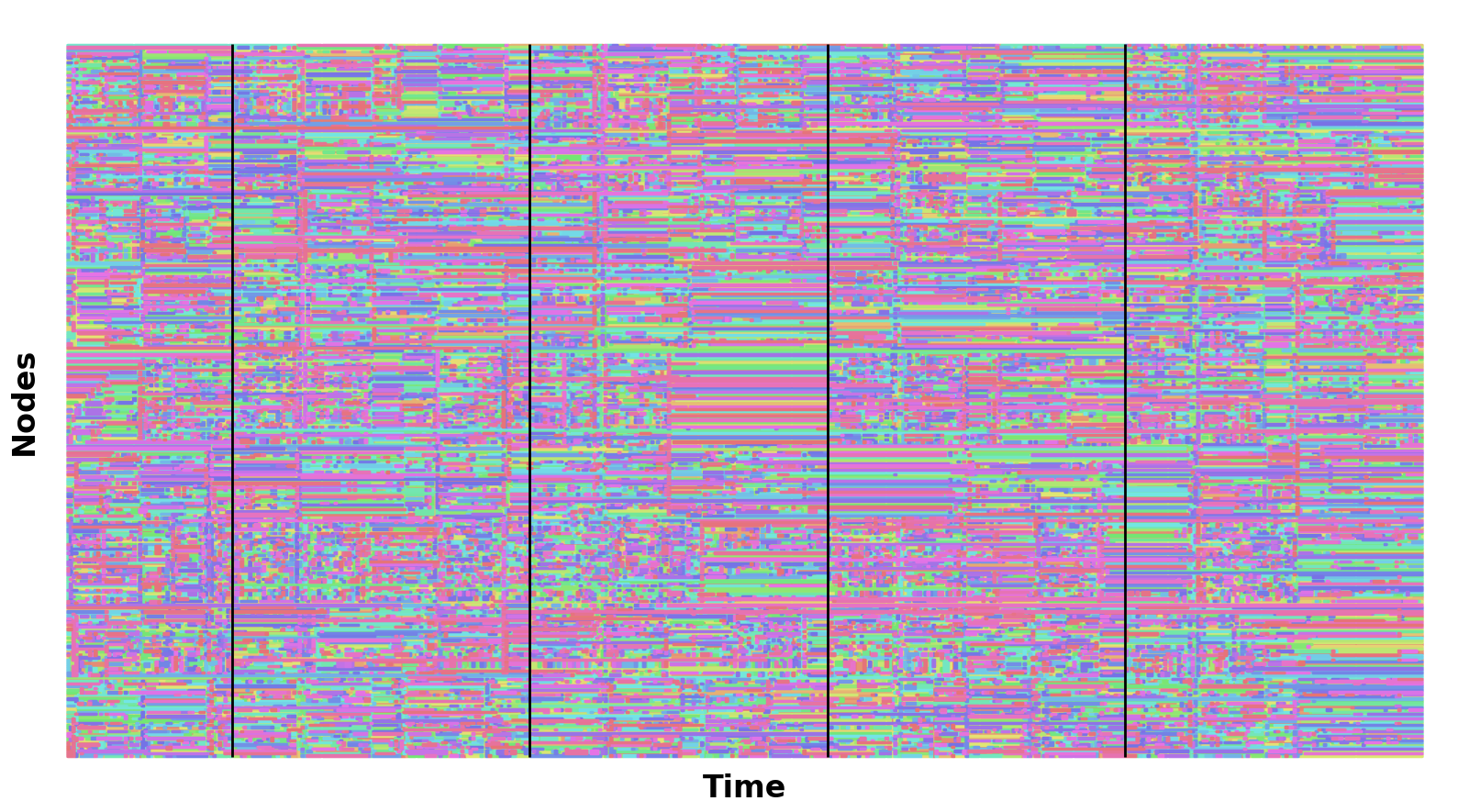}
         \captionsetup{width=0.95\textwidth}
         \caption{\fontsize{8}{9.5}\selectfont Worst structure for the mean membership expectation, with $23 564$ communities. Note the impact of the smoothness term: $Q_{\mathcal{L}, MM} = 0.98 - 2\rho$.}
     \label{fig:ns_worst}
     \end{subfigure}
\caption{Various dynamic community structures over the link stream of the Sociopatterns High School dataset. Vertical black lines represent nights during which interactions, and therefore communities, cease to exist.}
\label{fig:sociopatternhighchool_communities}
\end{figure}

In order to use MS-Modularity and NS-Modularity, we need to perform network aggregation into snapshots. We use several steps, from days to 5-minute aggregation step. Due to computational difficulties, we were not able to perform some steps, and could not go lower than 5 minutes for MS-Modularity. 
Given that the data are based on interactions occurring exclusively during school hours over a span of five days, we included nighttime periods in our experiment. Each night begins after the final interaction of the day and ends before the first interaction of the following day. During these nighttime periods, no communities are present.
Figures \ref{fig:sociopatternhighchool_heatmaps} and \ref{fig:sociopatternhighchool_communities} illustrates the results.

According to both L-Modularity versions, the MS-Modularity approach yields better community structures compared to the no-smoothing method. The advantage of MS-Modularity lies in its use of interslice weights, which facilitate the direct identification of continuous communities. In contrast, the no-smoothing method tends to overfit individual snapshots, and this overfitting is not adequately mitigated during the merging phase. This overfitting is exacerbated by more refined window aggregations.
Figures \ref{fig:ms_mm_heatmap} and \ref{fig:ms_jm_heatmap} illustrate the challenge of selecting appropriate window sizes and interslice weights for the MS-Modularity approach. According to L-Modularity, the finest window sizes tends to capture the most detailed temporal dynamics, and reasonably high interslice weights favor community continuity. However, determining the optimal combination of these parameters is complex. For instance, the best community structure for $Q_{\mathcal{C}, MM}$ (Fig. \ref{fig:ms_mm_best}) was revealed with the finest window aggregation of 5 minutes and an interslice weight ratio of 1, which is not the highest interslice weight tested. In contrast, the best community structure for $Q_{\mathcal{C}, JM}$ was revealed with the highest interslice weight ratio of 5, and a window aggregation of 30 minutes, not the finest one. Indeed, $Q_{\mathcal{C}, JM}$  tends to favor almost stationary communities with little or no evolution over time, whereas $Q_{\mathcal{C}, MM}$ allows for more variations in dynamic community structures. Furthermore, the best community structure for $Q_{\mathcal{C}, JM}$ was revealed with the NS method by aggregating data at the scale of a day and merging successive communities from each snapshot only if they share the exact same set of nodes (Fig. \ref{fig:ms_jm_best}). Both versions of L-Modularity can be used depending on specific analytical needs.

One could also consider the students' classes, provided in the dataset, as a \textit{ground truth} partition (Fig. \ref{fig:classes}). This partition yields a relatively high L-Modularity score, though not the highest. The experiment indicates that some students tend to interact beyond their class affiliations. Notably, Fig. \ref{fig:classes} and Fig. \ref{fig:ms_jm_best} share the same smoothness value,  as communities are considered to last the duration of each day, with only nights contributing to the smoothness term. However, Fig. \ref{fig:ms_jm_best} exhibits a higher L-Modularity score, demonstrating that at the daily scale, student communities do not strictly adhere to class boundaries.

Figure \ref{fig:ns_worst} illustrates the necessity of the smoothness term. It represents the best community structure according to L-Modularity when the smoothness term is discounted.

Note that we only compared L-Modularity scores of partitions obtained by methods that do not try to optimize it directly. The best partition discovered corresponds to something close to the expected ground truth given by classes, which shows that it does not fall into common traps such as favoring partitions overfitted at each timestep. However, one could expect to discover even better longitudinal communities, such as those distinguishing class periods from lunchtime and breaks between classes. This would require an algorithm able to better explore the space of possible partitions.

\section{Discussion}
\label{sec:discussion}
We presented a quality function for dynamic community structures in link streams that does not require a predefined time scale for analysis or any preprocessing of the natural temporal interactions. This function is capable of handling various configurations of dynamic community structures. Furthermore, like Modularity, L-Modularity is directly generalizable to directed, weighted, multigraphs, and bipartite temporal networks.

We believe that the approach we presented for generalizing Modularity can be seen as a first step in the direction of defining quality functions for link streams and that similar work could be done for other quality functions such as the Map Equation \cite{Rosvall_2009}. As a generalization of Modularity, one might expect similar limitations, such as the resolution limit problem \cite{Fortunato_2007}. These limitations should be analyzed in future work.

As it is defined here, L-Modularity cannot be used directly for community detection, because no efficient method exists to explore the space of possible temporal partitions of a link stream. A natural next step would consist in developing a Louvain-like algorithm to explore the space of possible partitions, searching to maximize L-Modularity. We anticipate that the results will provide valuable feedback on the behavior of L-Modularity with both synthetic and real data. Additionally, we hope this approach will uncover subtle dynamic community structures in real data, enhancing real-world data analysis.

\section*{Acknowledgements}
We would like to thank Matthieu Latapy for his valuable feedback and discussions. We also extend our gratitude to SAHAR for financing this project. Yasaman Asgari thanks the University of Zurich and the Digital Society Initiative for (partially) financing this project.

\section*{Materials availability}
The code necessary to reproduce the experiments and evaluate Longitudinal Modularity on various datasets is available here: \url{https://github.com/fondationsahar/dynamic_community_detection/}.

%
%
\bibliographystyle{splncs03}
\bibliography{refs}
\end{document}